\RequirePackage{lineno} 
\documentclass[aps,prl,twocolumn,groupedaddress,showpacs,floatfix,letterpaper]{revtex4}

\usepackage{graphicx}
\usepackage{multirow}
\usepackage{amsfonts}
\usepackage{amsmath}
\usepackage{amssymb}
\usepackage{hyperref}
\usepackage{color}
\usepackage{multirow}
\usepackage{lineno}

\def\al{\alpha}
\def\be{\beta}

\def\ta{\tau}

\def\De{\Delta}

\def\La{\Lambda}

\def\fr#1#2{\frac{#1}{#2}}

\def\ket#1{|{#1}\rangle}

\def\lsim{\mathrel{\rlap{\lower4pt\hbox{\hskip1pt$\sim$}}
    \raise1pt\hbox{$<$}}}
\def\gsim{\mathrel{\rlap{\lower4pt\hbox{\hskip1pt$\sim$}}
    \raise1pt\hbox{$>$}}}

\newcommand{\beq}{\begin{eqnarray}}
\newcommand{\eeq}{\end{eqnarray}}
\def\to{\rightarrow}

\def\no{\nonumber}

\def\maxf{10}
\def\minf{10}
\def\maxs{35}
\def\mins{2}
\def\maxa{3.6\times 10^{-26}}
\def\mina{6.3\times 10^{-28}}
\def\maxc{1.0\times 10^{-30}}
\def\minc{3.2\times 10^{-34}}

\def\alim{10^{-23}}

\def\clim{10^{-27}}

\begin{document}

\newcommand{\makespace}{\vspace{3 mm}}
\newcommand{\DP}{\displaystyle}

 \title{New Physics in Astrophysical Neutrino Flavor}

\date{\today}

\smallskip
\smallskip
\author{Carlos.~A.~Arg\"{u}elles$^{1,2}$,
Teppei Katori$^{3}$, and 
Jordi Salvado$^{1,2}$\\
}

\smallskip
\smallskip
\affiliation{
$^1$Department of Physics, University of Wisconsin, Madison, WI 53706, USA \\
$^2$Wisconsin IceCube Particle Astrophysics Center, Madison, WI 53706, USA \\
$^3$School of Physics and Astronomy, Queen Mary University of London, London, E1 4NS, UK \\
 }

\begin{abstract}
Astrophysical neutrinos are powerful tools to
investigate the fundamental properties of particle physics through their flavor content.
In this paper, we perform the first general new physics study on ultra high energy neutrino flavor content
by introducing effective operators.
We find that at the current limits on these operators,
new physics terms cause maximal effects on the flavor content, however,
the flavor content at Earth is confined to a region related to the assumed initial flavor content.
Furthermore, we conclude that a precise measure of the flavor content at Earth will provide orders of magnitude
improvement on new physics bounds.
Finally, we discuss the current best fits of flavor content of the IceCube data  and their interplay
with new physics scenarios. 
\end{abstract}

\maketitle



{\it Introduction} --- 
The existence of extra-terrestrial ultra high energy neutrinos
has been confirmed by the IceCube neutrino observatory~\cite{IC_UHEnu1,IC_UHEnu2}, opening
the possibility to study ultra high energy particle production mechanisms as well as new neutrino
physics \cite{Chung:1998zb,Feldstein:2013kka}.
The nature of these neutrinos from $\maxs$~TeV to $\mins$~PeV is still a puzzle, at the moment there are many
astrophysical and beyond the standard model candidate
sources~\cite{Bai:2013nga,Jordi_Sagittarius,Krauss:2014tna,Esmaili:2013gha,Anchordoqui:2014yva,Winter:2013cla}
that may produce these neutrinos. 
Currently, there is no statistically significant spatial correlation
between observed neutrinos and potential sources \cite{IC_Atmo2014,Taylor:2014hya}.


Even though the sources of these neutrinos remains unknown, 
it is still possible to find evidence of new physics. 
The vacuum neutrino propagation Hamiltonian is linearly proportional to the neutrino square mass differences and 
inversely proportional to the neutrino energy. 
For astrophysical ultra high energy neutrinos this operator is suppressed allowing to
look for extremely tiny new physics effects otherwise cannot be seen. 
In the standard oscillation scenario,
for any given initial flavor composition, the final composition, after the propagation,
lies in a small region on the flavor triangle close to $(\phi_e : \phi_\mu : \phi_\tau) = (1:1:1)$.
The flavor content of the astrophysical neutrinos has been studied
in~\cite{IC_UHEnu1,IC_UHEnu2,Mena:2014sja,Palomares-Ruiz:2015mka,IC_fratio,Palladino:2015zua,Kawanaka:2015qza}.
These analyses find flavor content is statistically consistent with the standard oscillations expectations.
Future data will clarify the IceCube astrophysical event flavor composition.

In this paper, we perform the first general new physics study of
the astrophysical neutrino flavor content by introducing effective operators
in the standard three neutrino scenario with unitary evolution.
This is the far most general approach to study new physics in astrophysical neutrino flavors,
and this approach covers many exotic particle physics models.
There are few cases we do not consider in this paper.
First, the model we work are limited within lepton number conservation,
and we do not consider models such as the neutrino-antineutrino oscillations~\cite{Kostelecky:2003cr,Diaz:2013iba}.
Second, we do not consider neutrino decay model which violates unitary evolution and was discussed elsewhere~\cite{Pagliaroli:2015rca}.
Similarly, we also do not consider models with sterile neutrino states~\cite{Conrad:2012qt}.
The sterile neutrino mixing matrix elements are known to be miniscule comparing with
the active neutrino mixing elements~\cite{Smirnov:2006bu,Kopp:2013vaa,Ignarra:2013mxa,Conrad:2012qt,Fukugita:1986hr},
and the contribution to the transition probability due to the sterile neutrinos
is suppressed by the sterile-active matrix element to the fourth power.


{\it Ultra High Energy Astrophysical Neutrino Oscillations} ---
Neutrinos change lepton flavors as they propagate macroscopic distances.
This is due to the fact that the neutrino propagation eigenstates are
not the eigenstates of the charged current weak interaction.
In presence of a dense medium the decoherent scattering interactions are important \cite{Delgado:2014kpa},
but in this paper we assume vacuum propagation.

In general the relation between the propagation eigenstates $\ket{\nu_i}$,
and the flavor eigenstates $\ket{\nu_\alpha}$, is given by a unitary transformation $V(E)$, 
\begin{equation}
  \ket{\nu_\alpha}= \sum_i V_{\alpha i}(E) \ket{\nu_i}~. 
\end{equation}
For astrophysical neutrinos, 
the propagation distance is much longer than the oscillation length, 
and in this limit the oscillation 
from flavor state $\ket{\nu_\alpha}$ to a flavor state $\ket{\nu_\beta}$ can be averaged,
\beq
\bar P_{\nu_{\al} \to \nu_{\be}}(E)&=&
\sum_{i}\left|V_{\al i}(E)\right|^2\left|V_{\be i}(E)\right|^2~,
\label{eq:osc_incoh}
\eeq
where the probability depends only on the mixing matrix elements $|V_{\al i}(E)|$,
which is in general energy dependent.
Using the probability given in this equation 
and the flux at production $\phi^p_\alpha$,
we can calculate the neutrino flux at Earth, $\phi^{\oplus}_\beta (E)$, for a flavor $\beta$. 
It is more convenient to define the energy averaged flavor composition as 
\beq
\bar \phi^{\oplus}_\beta= \frac{1}{|\Delta E|}\int_{\Delta E} \sum_\alpha \bar P_{\nu_{\al} \to \nu_{\be}}(E) \phi^p_\alpha(E)  dE~,
\eeq
where we assume $E^{-2}$ power law for the production flux
and $\Delta E=$[$\maxf~\mathrm{TeV}$, $\minf~\mathrm{PeV}$].
Note, however, that our main results are largely insensitive to the spectral index.
We also assume that all flavors have the same energy dependence at the source.


In astrophysics charged pion decay from proton-proton collisions is one of the preferred neutrino production channels.
In this scenario the initial flavor composition is $(\phi_e:\phi_\mu:\phi_\tau)=(1:2:0)$.
Other scenarios such as rapid muon energy loss produce $(0:1:0)$, neutron decay dominated
sources produce $(1:0:0)$ are of interest, while compositions such as $(0:0:1)$ are not
expected in the standard particle astrophysics scenarios. 
In order to plot the flavor content in a flavor triangle we introduce 
the flavor fraction, $\alpha^{\oplus}_\beta=\bar\phi^{\oplus}_\beta / \sum_\gamma \bar\phi^{\oplus}_\gamma$.

For the vacuum propagation, 
the Hamiltonian of the standard neutrino oscillation
only depends on the neutrino mass term,
\begin{equation}
H= \frac{1}{2E}  U 
\left(\begin{array}{ccc}
0  & 0 & 0 \\
0 & \De m^2_{21}& 0 \\ 
0 & 0 & \De m^2_{31}
\end{array}\right)
U^{\dagger}=\frac{1}{2E}U M^2 U^\dagger
~,
\end{equation}
where $E$ is neutrino energy, $\Delta m_{ij}^2=m_i^2-m_j^2$,
and $U$ is the standard lepton mixing matrix $U$. 
Throughout this paper we assume the normal mass ordering.
We also performed the same study by assuming the inverted mass ordering,
however, differences are minor and mass ordering does not affect any of our main conclusions.

\begin{figure}[t!]
  \begin{center}
    \includegraphics[width=\columnwidth]{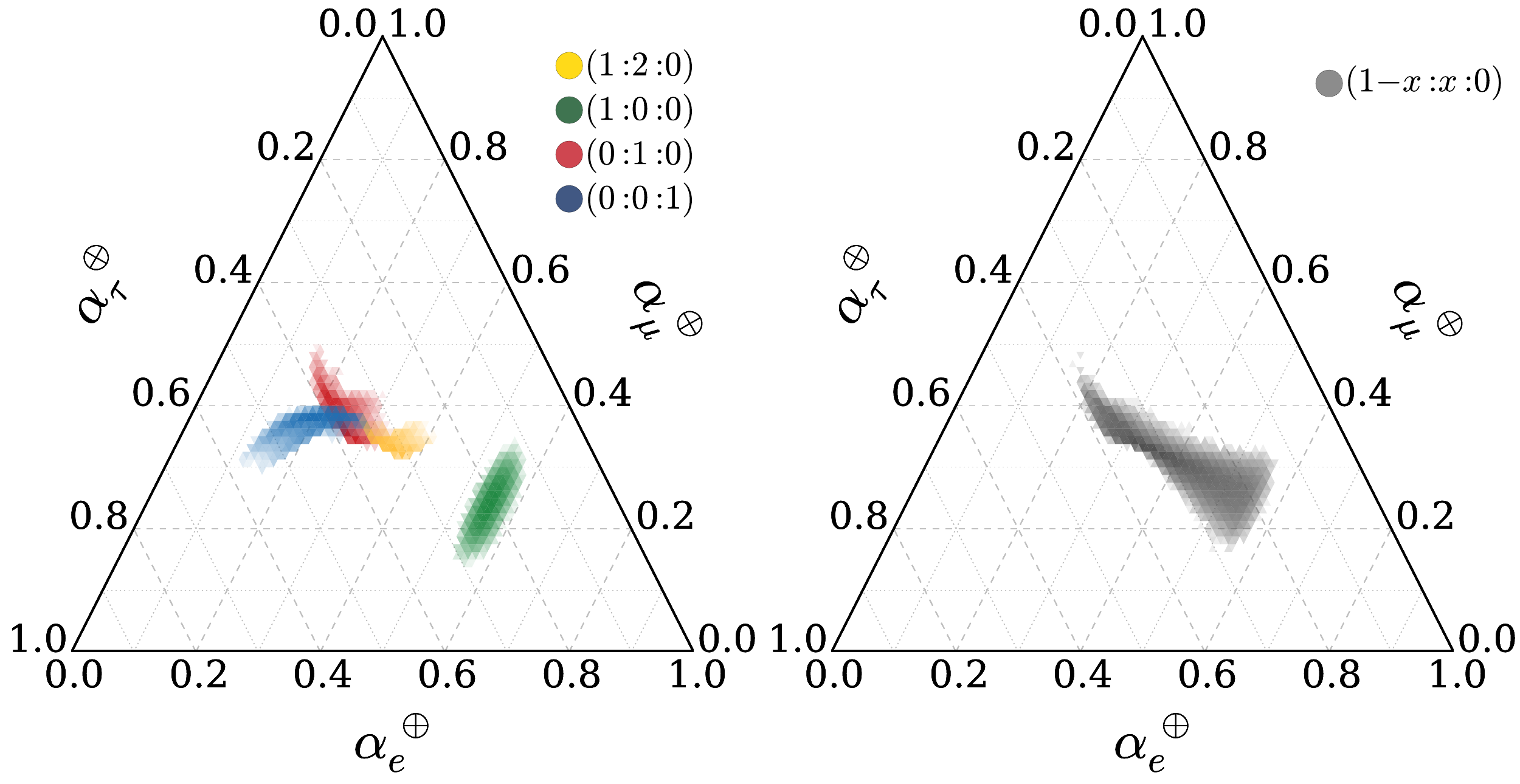}
  \end{center}
\vspace{-2mm}
\caption{Allowed regions of the flavor content at Earth using
  the priors on the mixing angles and errors given from the current neutrino oscillation measurements.
  In the left plot, the different colors correspond to different assumptions on flavor content at the production.
  The color intensity is proportional to the probability density.
  In the right plot, we further sample the initial flavor content as $(x:1-x:0)$.
  }
\label{fig:vacuum}
\end{figure}

The current measurements of the standard neutrino 
oscillation experiments allows us to determine
the astrophysical neutrino flavor content at detection given an assumption of the neutrino production.
In Fig.~\ref{fig:vacuum} we show allowed regions of the flavor content at Earth,
where we use the standard mixing angles and their errors from the global fits~\cite{Gonzalez-Garcia:2014yaa} in
order to produce probability density distributions for the flavor content.
Since the CP-phase is not strongly constrained by neither terrestrial~\cite{Abe:2013hdq,Adamson:2013ue} 
nor astrophysical~\cite{Chatterjee:2013tza} neutrinos, we assume a flat distribution from 0 to $2\pi$.
Note that for simplicity we use the larger of the asymmetric errors and implement them as Gaussian. 
In the left plot, we assume four different production flavor composition hypotheses.
We observe that all the allowed regions of astrophysical neutrino flavor content at Earth are close to $(1:1:1)$,
except when the initial flavor content is $(1:0:0)$~\cite{Palladino:2015vna}. 
In the right plot, we show the allowed region of the flavor content of the astrophysical neutrinos
with all possible astrophysical production mechanisms,
{\it i.e.}, the production flavor composition is sampled with $(x:1-x:0)$ uniformly on $x$~\cite{Bustamante:2015waa}.
Therefore, this rather narrow band covers all possible scenarios of the standard neutrino oscillations
with the standard astrophysical neutrino production mechanisms.

{\it New Physics in Effective Hamiltonians} --- 
An effective way of introducing new physics in neutrino oscillations is by introducing new operators.
The full Hamiltonian that incorporates the new physics operators, in the flavor basis, can be expressed as,
\begin{equation}
  H=\frac{1}{2E}U M^2 U^\dagger +\sum_n \left(\frac{E}{\Lambda_n}\right)^n \tilde U_n O_n \tilde U_n^\dagger
  = V^\dagger(E) \Delta V(E)~,\no 
\end{equation}
where
$O_n=\mathrm{diag}(O_{n,1},O_{n,2} ,O_{n,3})$
and
$\Delta=\mathrm{diag}(\Delta_{1},\Delta_{2} ,\Delta_{3})$.
$O_n$ and $\La_n$ set the scale of the new physics
and $\tilde U_n$ is the mixing matrix that describes the new physics flavor structure.
In the effective theory approach, lower order operators are more relevant, 
thus in this work we will only study the first terms in the expansion, namely $n=0$ and $n=1$. 

Although in this work we will study $n=0$ and $n=1$,
results can be extended to higher orders. 
These new operators can be interpreted in different new physics contexts.
Some examples for $n=0$ new physics are couplings between neutrinos and space time torsion~\cite{DeSabbata:1981ek},
CPT-odd Lorenz violation~\cite{Mewes_1,Mewes_HighOrderNu,Barger:2000iv,Coleman:1998ti},
and non-standard neutrino interactions~\cite{Valle:1987gv,Guzzo:1991hi,Grossman:1995wx,Bergmann:1999rz}.
As for $n=1$ new physics operators, CPT-even Lorentz violation~\cite{Glashow:1997gx,Diaz_UHE} and equivalence principle
violation~\cite{Gasperini:1989rt,Butler:1993wi} are possible examples.

There are some constraints from neutrino oscillation experiments to these effective
operators in the context of Lorentz and CPT violation~\cite{Kostelecky:2008ts}.
The most stringent limits on certain parameters are obtained from Super-Kamiokande
and IceCube atmospheric neutrino analyses~\cite{SK_LV,IC_LV}.
In this context the CPT-odd and CPT-even Lorentz violation coefficients
are constrained to be $\sim\alim~\mathrm{GeV}$ and $\sim\clim$
depending on the flavor structure $\tilde U_n$.
These constraints can be used to set the scales of $n=0$ and $n=1$ operators introduced in this paper. 
For example, we set $O_0=1\times\alim~\mathrm{GeV}$ as a current limit of the $n=0$ operator,
and $O_1=1\times\alim~\mathrm{GeV}$ with $\La_1=1~\mathrm{TeV}$ as a current limit of $n=1$ operators,
where $\fr{O_1}{\La_1}=\clim$.
Through this paper we have assumed the scale of $O_1$ is of
the order of $O_0$ without loss of generality. 

{\it Anarchic Sampling prediction and IceCube Results} --- 
In order to predict the flavor composition at Earth in the presence of new physics,
the values of the mixing matrices $\tilde U_n$ should be specified.
In order to show a prediction with new physics operators,
we have to account for all the free parameters in the mixing matrix;
we use a random sampling scheme to construct the mixing matrix.
A well established schema is the anarchic
sampling~\cite{Murayama_anarchy1,Murayama_anarchy2,deGouvea:2012ac,deGouvea:2003xe},
which samples a flat distribution given by the Haar measure, 
\begin{equation}
  d\tilde U_n = d\tilde s^2_{12} \wedge d\tilde c^4_{13} \wedge d\tilde s^2_{23} \wedge d\tilde\delta~,
\end{equation}
where, $\tilde s_{ij}$, $\tilde c_{ij}$, and $\tilde\delta$ correspond to sines, cosines, 
and phase for the new physics $n$-operator mixing angles. 
We omit the Majorana phases since they do not affect neutrino oscillations. 


\begin{figure}[t!]
  \begin{center}
  \includegraphics[width=\columnwidth]{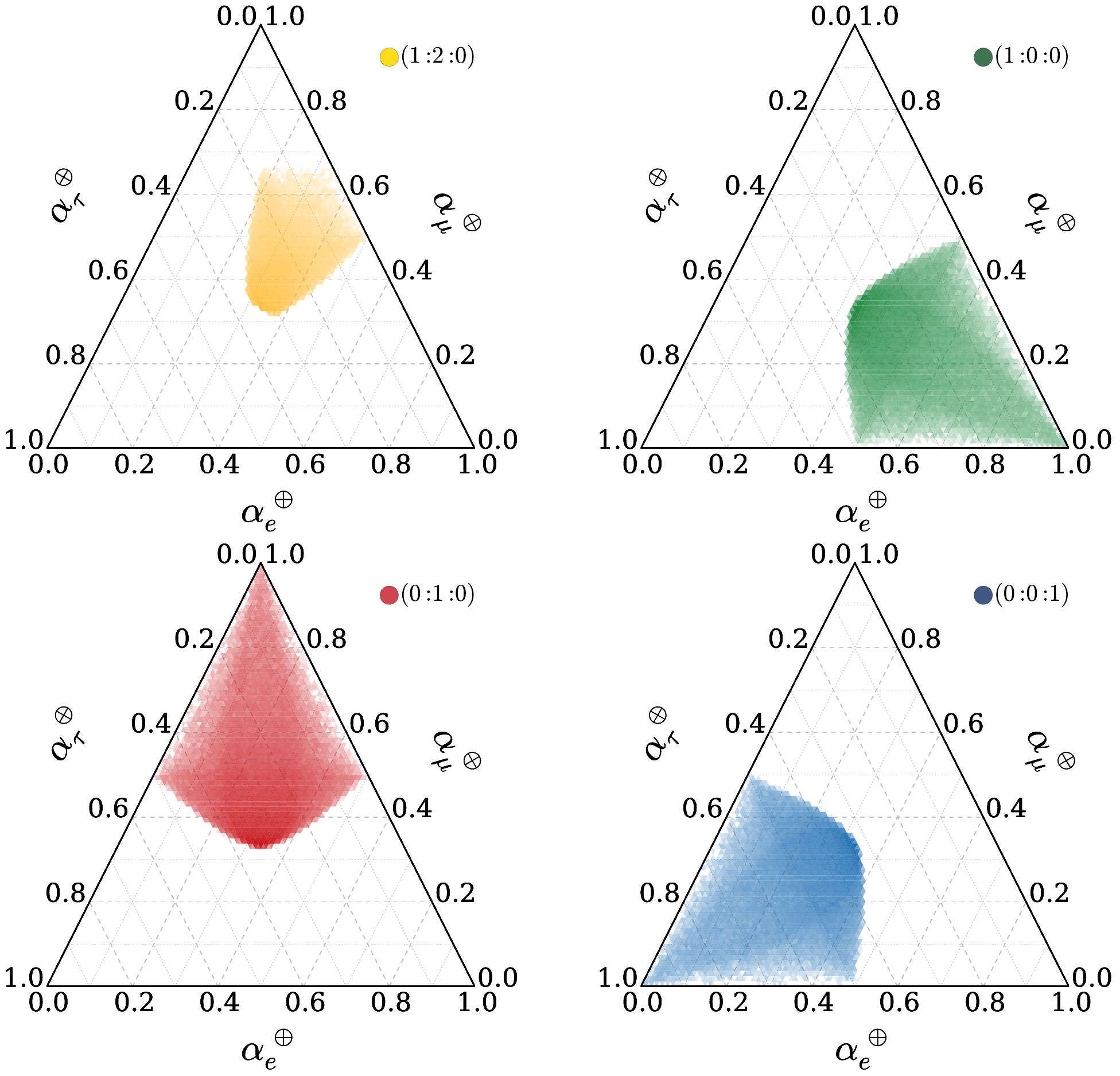}
  \vspace{-2mm}
  \end{center}
\caption{
  Allowed region using anarchic sampling on the mixing angles for the new physics operator
  when the mass term in the Hamiltonian is neglected.
  The different plots correspond to different assumption on flavor content at production.
  The color intensity is proportional to the probability predicted by anarchic sampling.
}
\label{fig:a23a33}
\end{figure}

In Fig.\ref{fig:a23a33} we show the allowed regions using anarchic sampling in the
case where $H=\left(\frac{E}{\Lambda_n}\right)^n \tilde U_n O_n \tilde U^\dagger_n$.
In this case, we neglect the mass term and we are considering that the Hamiltonian has only
one operator, {\it i.e.}, $V=\tilde U_n$, and the result does not depend on $n$.
Each plot in this figure correspond to a different production flavor composition. We show the pion decay
production $(1:2:0)$ [yellow], beta decay $(1:0:0)$ [green], muon cooling $(0:1:0)$ [red] and for
completeness we show the exotic $\nu_\ta$ dominant model $(0:0:1)$ [blue].
The color density in these plots is a representation of the probability given by the anarchic sampling. 

\begin{figure}[t!]
  \begin{center}
    \includegraphics[width=\columnwidth]{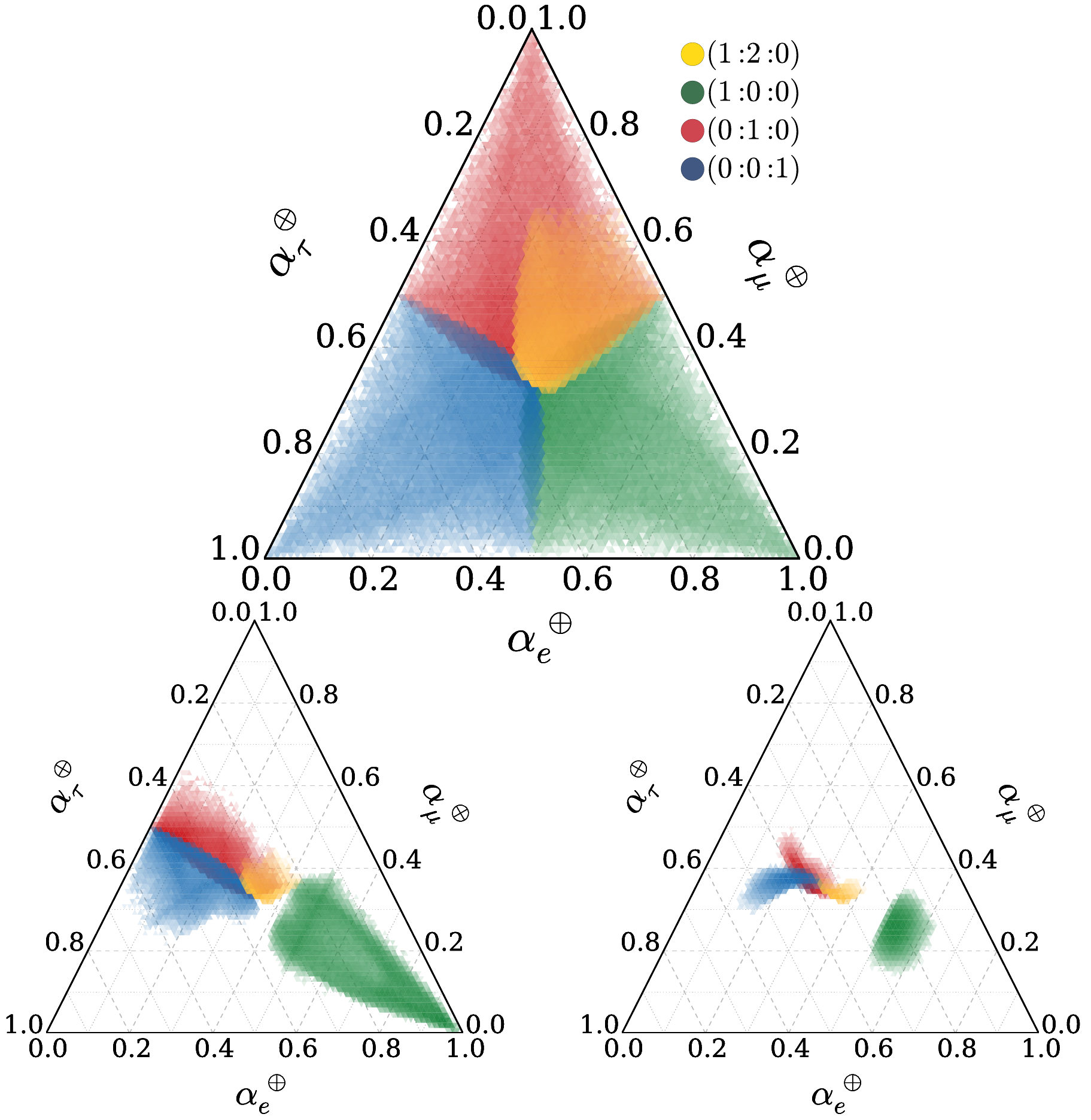}
    \end{center}
\vspace{-2mm}
\caption{
  Allowed region using anarchic sampling on the mixing angles for the new physics $n=0$ operators.
  The top plot corresponds to the current limits on $n=0$ operator;
  the bottom left plot corresponds to $O_{0}=\maxa$~GeV,
  while the bottom right plot corresponds to $O_{0}=\mina$~GeV.
}
\label{fig:aterm}
\end{figure}

In Fig.\ref{fig:aterm} we show the case where we have a mass term and the $n=0$ operators.
In the top plot, we set $O_{0}=1.0\times\alim$~GeV, corresponding to the order of
the current best limit on this operator. 
On the bottom left plot we set $O_{0}=\maxa$~GeV and
the bottom right plot we set $O_{0}=\mina$~GeV. 
These values are chosen because they have the same magnitude as the mass term
with neutrino energy of $E_\nu=\maxs$~TeV  and $E_\nu=\mins$~PeV respectively.
In this plot, the colors represent different assumptions in the production flavor content,
and the color intensity is the probability given by the anarchic sampling as in Fig.\ref{fig:a23a33}.

\begin{figure}[t!]
  \begin{center}
    \includegraphics[width=\columnwidth]{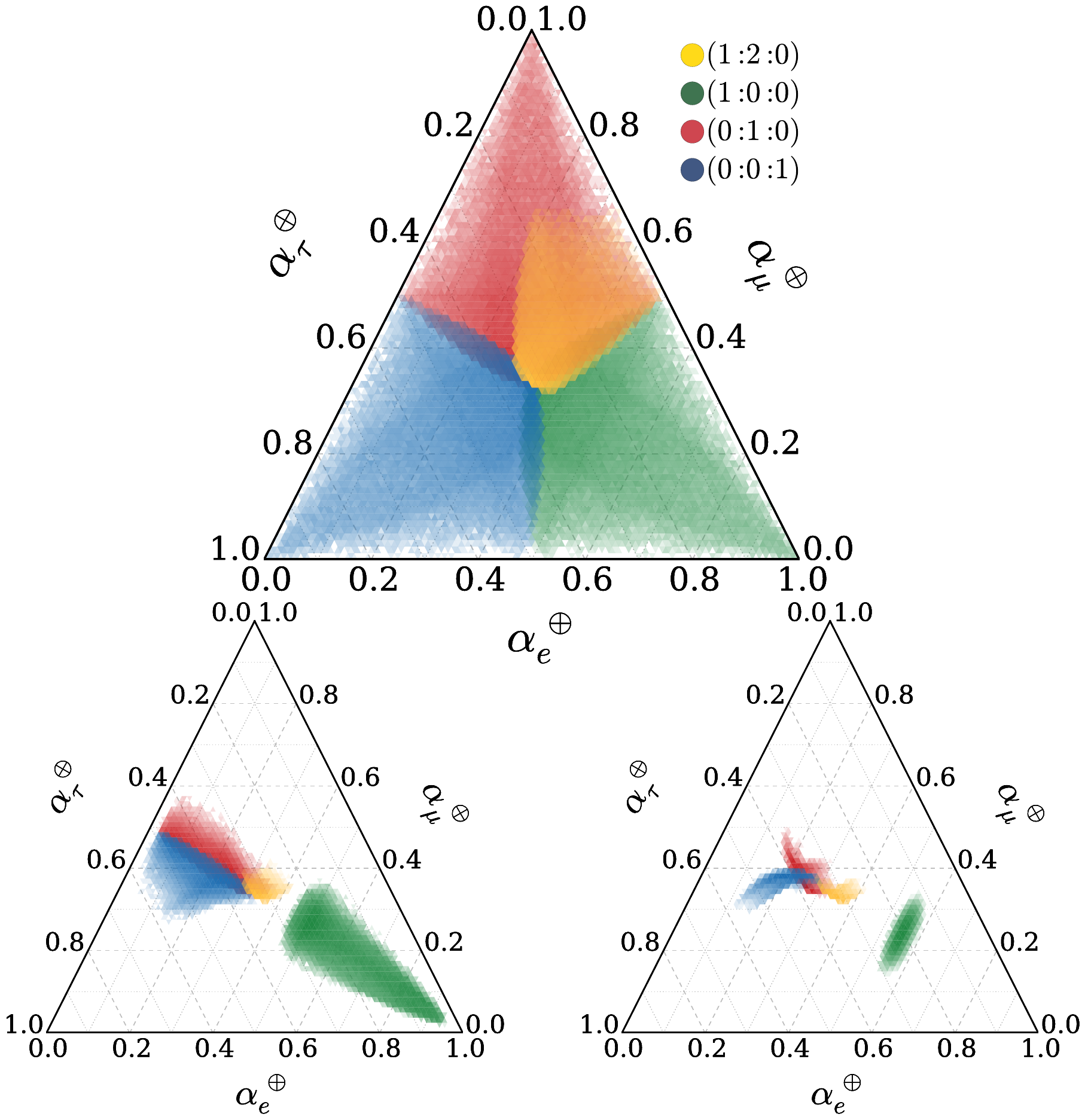}
    \end{center}
\vspace{-2mm}
\caption{
  Allowed region using anarchic sampling on the mixing angles for the new physics $n=1$ operators.
  The top plot corresponds to the current limits on $n=1$ operator;
  the bottom left plot corresponds to  $O_{1}=\maxa$~GeV and $\Lambda_1=\maxs$~TeV ($\fr{O_1}{\La_1}=\maxc$),
  while the bottom right plot corresponds to $O_{1}=\mina$~GeV and $\Lambda_1=\mins$~PeV ($\fr{O_1}{\La_1}=\minc$).
}
\label{fig:cterm}
\end{figure}

In Fig.\ref{fig:cterm} we show the case for the $n=1$ operators.
The color notations and their intensities have equivalent meaning as Fig.\ref{fig:aterm}.
As before, the top plot we set the new physics operator to the current best limit
$\fr{O_{1}}{\La_1}\sim\clim$. 
This is achieved by choosing $O_{1}=O_{0}=1.0\times\alim$~GeV and $\Lambda_1=1$~TeV.
In the bottom left plot, $O_{1}=\maxa$~GeV and $\Lambda_1=\maxs$~TeV are used,
and in the bottom right plot the parameters are $O_{1}=\mina$~GeV and $\Lambda_1=\mins$~PeV.
These choices make new physics to be the same magnitude as the mass term 
with neutrino energy of $E_\nu=\maxs$~TeV and $E_\nu=\mins$~PeV, respectively.
In other words, these choices explore new physics down to $\fr{O_1}{\La_1}=\maxc$
and $\fr{O_1}{\La_1}=\minc$.
This can be compared, for example, to the aforementioned best limits of Lorentz
and CPT violation in neutrino sector~\cite{IC_LV,SK_LV}.
The potential limits from astrophysical neutrino flavor content can
be well beyond what terrestrial neutrino experiments can achieve.

From Fig.\ref{fig:aterm} and Fig.\ref{fig:cterm}
we observe that the allowed regions in the flavor triangle change in a similar way as a function 
of the energy scale.
This is true for any higher operators,
because what matters is the scale where they dominates over standard neutrino mass term,
and these two operators are sufficient to predict behaviors of any higher order operators. 
Comparing Fig.\ref{fig:aterm} and Fig.\ref{fig:cterm} with respect to Fig.\ref{fig:a23a33} where the allowed regions are more symmetric, 
there is a preferred region along the vacuum oscillation triangle shown in Fig.\ref{fig:vacuum}.
It is interesting to notice that due to the unitary evolution and the fact that the oscillations are averaged,
for a given production flavor content, only a subset of the flavor triangle is accessible.
The pion decay production mechanism $(1:2:0)$ is one of the most natural
astrophysical scenarios for high energy neutrino production.
From Fig.\ref{fig:aterm} and Fig.\ref{fig:cterm} the allowed region for this case is the smallest,
which means that if future measurements exclude this region,
the pion production dominant mechanism is excluded regardless of the presence of new oscillation physics.

In the analyses of the IceCube high energy neutrino events, different results have been shown.
The first result~\cite{Palomares-Ruiz:2014zra}, using the IceCube result~\cite{IC_UHEnu2},
showed a best fit at $(1:0:0)$ disfavoring $(1:1:1)$ at 92\%~C.L.
Later, the same authors did an improved analysis~\cite{Palomares-Ruiz:2015mka} including energy dependence and
extra systematic errors, finding that the best fit may move considerably depending on the
features of the energy spectrum such as including an energy cutoff or not.
The IceCube collaboration later published an analysis of the flavor ratio
above $30$~TeV \cite{IC_fratio} finding a best fit at $(0:\fr{1}{5}:\fr{4}{5})$, as well as excluding
$(1:0:0)$ and $(0:1:0)$ at more than 90\%~C.L.
This IceCube result shows a best fit dominated by the $\nu_\tau$ component, which
can be explained by the correlation between the energy cutoff and the Glashow resonance,
as noted by ~\cite{Palomares-Ruiz:2015mka}.
In obtaining this best fit, the IceCube collaboration has assumed an equal amount of neutrinos and antineutrinos,
which best corresponds to a proton-proton source.
On the other hand, if the neutrino source is proton-photon dominated then the neutrino-antineutrino ratio
weaken making the previous conclusion.
It is interesting to notice that if this IceCube best fit does not change considerably after adding more data,
the production mechanism has to include a $\nu_\tau$ component.
This is because the new physics in the propagation can not give the best fit value for any plausible astrophysical scenarios.
This implies not only new physics in the neutrino oscillations, but also new physics in the production mechanism.

{\it Conclusions} ---
We performed the first new physics study on the astrophysical neutrino flavor content using effective operators
in the standard three neutrino scenario.
These operators can represent a variety of models such as Lorentz and CPT violation,
violation of equivalent principle, cosmic torsion, non-standard interactions, etc,
making this work to be the most general study of new physics in astrophysical neutrino flavor content to date.

We found that large effects in the flavor content at Earth are still allowed with given terrestrial bounds
on new physics in the neutrino sector.
This implies that an accurate measurement of the flavor content will provide stronger bounds on new physics.
Furthermore, there are regions on the flavor triangle that cannot be accessed even in the presence of new physics
in the neutrino oscillations for any of the plausible astrophysical mechanisms.
Interestingly, that the most natural astrophysical mechanism, pion decay,
has the smallest region in the flavor triangle even when new physics is considered.
The real astrophysical neutrino production mechanism in nature may be the combination of channels, but our results hold for such a case.
Therefore, a higher statistics measurement by future neutrino telescopes, such as IceCube-Gen2~\cite{Aartsen:2014njl},
could reveal not only the initial neutrino flavor ratios, but also the presence of new physics in neutrinos. 


{\it Acknowledgments} --- 
We thank Logan Wille, Markus Ahlers, and Jorge D\'{i}az for useful discussions.
The authors acknowledge support from the Wisconsin IceCube Particle Astrophysics Center (WIPAC).
C.A. and J.S. were supported in part by the National Science Foundation (OPP-0236449
and PHY-0969061) and by the University of Wisconsin Research Committee with funds
granted by the Wisconsin Alumni Research Foundation. 

\bibliographystyle{apsrev}
\bibliography{fratio}

\end{document}